\documentstyle[aps,epsf,epsfig,floats]{revtex}
\begin{document}
\preprint{\vbox{\hbox{DOE/ER/40762-232}\hbox{UM PP\#01-057}}}
\title{Model-Independent Predictions for Low Energy Isoscalar Heavy Baryon 
Observables in the Combined Heavy Quark and Large $N_c$ Expansion.}
\author{Z.~Aziza Baccouche, Chi--Keung Chow, Thomas D.~Cohen and Boris
A.~Gelman}
\address{Department of Physics, University of Maryland, College Park, MD 
20742-4111.} 
\maketitle
\begin{abstract}
Model-independent predictions for excitation energies, semileptonic
form factors and electromagnetic decay rates of isoscalar heavy baryons
and their low energy excited states are discussed in terms of the
combined heavy quark and large $N_c$ expansion. At leading order, the 
observables are completely determined in terms of the known excitation energy 
of the first excited state of $\Lambda_c$. At next-to-leading order in
the combined expansion all heavy baryon observables can be expressed in a 
model-independent way in terms of two experimentally measurable quantities.
We list predictions at leading and next-to-leading order.
\end{abstract}



A number of nonperturbative expansions have been used to study low-energy
hadronic observables. The heavy quark expansion has been very successful in
describing hadrons containing one heavy quark. In the heavy quark limit 
the heavy quark spin-flavor decouples from the dynamics and hadronic spectrum
exhibits an approximate heavy quark spin-flavor symmetry. Heavy quark
effective theory (HQET) is a consistent framework to study the $1/m_Q$ 
corrections to the heavy quark limit \cite{HQ1,HQ2,HQ3,HQ4,HQ5,HQ6}. Another
useful limit is the large $N_c$ limit where the number of colors $N_c$ of
the QCD gauge group is taken to be large \cite{LN1,LN2}. In the large $N_c$ 
limit the baryon sector of QCD exhibits an approximate light quark 
spin-flavor symmetry; {\it eg.} as $N_c$ goes to infinity the nucleon and 
$\Delta$ become degenerate \cite{SF1,SF2,SF3,SF4,SF5}. For heavy baryons, 
{\it i.e.} baryons containing one (c or b) heavy quark it is useful to 
consider a double limit and a combined heavy quark and large $N_c$ expansion
\cite{hb0}. In Ref.~\cite{hb1,hb2,hb3} isoscalar heavy baryons have been 
considered near the combined limit.

The combined heavy quark and large $N_c$ expansion is formulated in terms of 
the counting parameter $\lambda \sim \Lambda_{had}/m_Q, \, 1/N_c$ with 
$\Lambda_{had} \, N_c/m_Q$ arbitrary (here $\Lambda_{had}$ is a 
typical hadronic scale). As $\lambda$ approaches zero, the heavy 
baryon spectrum exhibits an approximate dynamical symmetry---a contracted 
$O(8)$ symmetry---which connects orbitally excited states of the heavy baryons
to the ground state \cite{hb0,hb1}. An effective theory describing low-energy
excitations of heavy baryons can be developed using the counting rules
consistent with a contracted $O(8)$ symmetry \cite{hb2}. The purpose of the 
present paper is to use the formalism of Ref.~\cite{hb3} to derive 
model-independent relations among observables characterizing isoscalar heavy
baryons and their decays. At present, experimental data is available only for
the excitation energies of the doublet of the first excited state of
$\Lambda_c$. At leading order in the combined expansion this energy completely
determines the excitation energy of the low-energy orbitally excited state
of $\Lambda_b$ as well as the dominant semileptonic form factors and 
electromagnetic decay rates. Such predictions are necessarily rather crude as 
our expansion is in powers of $\lambda^{1/2}$. At next-to-leading order an 
additional 
observable is required to make further model independent predictions. As data
for more heavy baryon observables (such as the excitation energy of $\Lambda_b$
baryon, semileptonic and radiative decay rates of $\Lambda_c$ and $\Lambda_b$
baryons and their excited states) become available, the next-to-leading order
predictions can be tested against experimental results. In this paper
we discuss which observables are most sensitive to the NLO corrections, and,
thus, can be used to extract two model-independent constants which arise at 
leading and next-to-leading order in the effective expansion. Using the 
formalism of Ref.~\cite{hb3} we derive model independent relations among a 
number of observables characterizing isoscalar heavy baryons and their decays.

The effective degrees of freedom near the combined limit are the collective
motion of the brown muck---the light quarks and gluons---relative to the
heavy quark. As shown in Ref.~\cite{hb2}, the energy of the collective
excitations is of order $\lambda^{1/2}$ while the energy of the intrinsic 
excitations of the 
brown muck itself are of order $\lambda^0$. The effective operators that 
excite the collective degrees of freedom can be defined from the corresponding
QCD operators in the combined limit. The construction of these operators from 
QCD in a model-independent way was discussed in detail in Ref.~\cite{hb1,hb2}. 
They are the total momentum $\vec P$ (a constant of the motion) and its
conjugate position operator $\vec X$ of the entire system. These 
operators commute up to corrections of order $\lambda$ with another pair of 
the collective conjugate operators $\vec p$ and $\vec x$ which are related to
the momentum and position of the brown muck relative to heavy quark. 

We have so far focused on the isoscalar heavy baryons, $\Lambda_c$ and
$\Lambda_b$, and their excited states. The effective Hamiltonian for these
states can be written solely in terms of the collective operators discussed
above. As shown in Ref.~\cite{hb1}, the contracted $O(8)$ symmetry
naturally arises in the combined limit if the counting rules of the
collective operators are given by,
\begin{eqnarray}
(x, \, X, \, X_Q, \, X_\ell) \, & \sim & \, \lambda^{1/4} \, ,
\nonumber \\
(p, \,\,  P, \,\, P_Q, \,\, P_\ell) \, & \sim & \, \lambda^{-1/4} \, ,
\label{CR}
\end{eqnarray}
in the sense that the typical matrix elements of these operators between the
low-lying heavy baryon states scale as $\lambda^{1/4}$ and $\lambda^{-1/4}$.
The effective Hamiltonian based on these counting rules up to terms
of order $\lambda$ was derived in Ref.~\cite{hb2}:
\begin{equation}
\begin{array}{ccccccccccccccc}
 H_{\rm eff} & =  & ( m_H \, + \, m_N )& \, + \, & {c}_{0} &\,+
\, & \left (\frac{P^2}{2 (m_N + m_H)}+ \frac{p^2}{2 \mu_Q}  \, + 
\, \frac{1}{2} \kappa x^2 \right ) & \, + \, &
\frac{1}{4!} \alpha x^4 & \,  + \, & {\cal O}(\lambda^{3/2}) \, , \\  & &
\parallel & & \parallel & & \parallel & & \parallel & & \\  
 & & {\cal H}_{\lambda^{-1}} & & {\cal H}_{\lambda^{0}} 
&  & {\cal H}_{\lambda^{1/2}} & & {\cal H}_{\lambda^{1}} & &
\end{array}
\label{HEFF}
\end{equation}
where $m_N$ is the nucleon mass; $m_H$ is the mass of the heavy meson 
containing the same heavy quark as the heavy baryon. These masses are 
measurable quantities while the corresponding brown muck and the heavy quark 
masses are not. In the combined limit, the difference between the
heavy quark and the corresponding heavy meson mass is of order $\lambda^0$;
the difference between the brown muck mass and the nucleon mass is of order
$\lambda^{0}$ as well. The heavy meson and the nucleon masses scale as 
$\lambda^{-1}$ in the combined limit \cite{hb2}. The reduced mass, $\mu_Q$,
which is of order $\lambda^{-1}$, is given by,
\begin{equation} 
\mu_Q \, = \, \frac{m_N  m_H}{m_N + m_H} \, .
\label{mu}
\end{equation}
In eq.~(\ref{HEFF}), ${\cal H}_{\lambda^{n}}$ refers to the piece of the 
Hamiltonian whose contribution is of order $\lambda^{n}$. The term
${\cal H}_{\lambda^{1/2}}$ is referred to as leading order (LO), and 
${\cal H}_{\lambda}$ as next-to-leading order (NLO). The combined expansion
is in powers of $\lambda^{1/2}$.

The effective Hamiltonian in eq.~(\ref{HEFF}) contains three phenomenological
parameters---$c_0$, $\kappa$ and $\alpha$ which are of order $\lambda^0$. 
At order $\lambda^0$ these constants are independent of the heavy quark 
flavor. However, they contain $1/m_Q \sim \lambda$ corrections. As seen from 
eq.~(\ref{HEFF}), the $1/m_Q$ corrections to $\kappa$ and $\alpha$ contribute
only at order $\lambda^{3/2}$, {\it i.e.} at next-to-next-to-leading order
(NNLO) and higher. Hence they can be neglected when we work only up to NLO.
The $1/m_Q$ correction to $c_0$ contributes at order $\lambda$, {\it i.e.}
at NLO. However, the dynamics of the collective degrees of freedom does not
depend on the constant $c_0$ which only determines overall ground state energy.
Thus, only two flavor-independent constants up to NLO determine the low-energy
dynamics of the effective degrees of freedom---the collective motion of
the brown muck relative to the heavy quark. They can be extracted from data in
a model-independent way provided at least two relevant observables are 
measured.

The phenomenology of the $\Lambda_c$ and $\Lambda_b$ baryons and their 
first orbitally excited states based on the effective Hamiltonian in 
eq.~(\ref{HEFF}) was discussed in Ref.~\cite{hb3}. In addition to the 
Hamiltonian, the effective operators determining semileptonic form factors and 
electromagnetic decay rates have been derived in a model-independent way 
in the combined expansion \cite{hb2}. Remarkably up to NLO, these effective 
operators do not contain additional phenomenological parameters. In what 
follows, we present the model-independent predictions of the effective theory 
at LO and NLO. At LO, these predictions are the same as obtained from the 
bound state picture of the heavy baryon in which the heavy baryon is thought 
of as the bound state of the heavy meson and an ordinary baryon ({\it eg.} 
the nucleon) \cite{bst1,bst2,bst3,bst4,bst5,bst6,bst7,bst8,bst9,HQ15}. 
However, the formalism of Ref.~\cite{hb3} puts these results in a 
model-independent framework. 

A useful eigenbasis for the low-lying heavy baryons is spanned by 
the eigenstates of the leading order terms of the effective Hamiltonian,
{\it i.e.} $|\Lambda_Q; N, l, m; \sigma \rangle$, where $N, l, m$ are 
the quantum numbers of the three-dimensional harmonic oscillator and $\sigma$ 
is the third component of the heavy quark spin ($\sigma= \pm 1/2$). The heavy
quark spin decouples from the dynamics up to NNLO. As a result, the
total angular momentum of the heavy baryon is $J=l + s_Q=|l \pm 1/2|$. The
heavy baryons are the eigenstates of the total angular momentum, {\it i.e.}
they are appropriate linear superpositions of the states 
$|\Lambda_Q; N, l, m; \sigma \rangle$. We denote these states by
$|\Lambda_Q; N, J, J_z \rangle$. The ground states in both the charm and bottom
sectors, $\Lambda_c$ and $\Lambda_b$, have been observed. In addition,
the doublet of the first orbitally excited state---($\Lambda_{c1}$, 
$\Lambda_{c1}^{*}$)---has been observed \cite{exp1,exp2,exp3}. For further
convenience we use the following notation for these states:
\begin{eqnarray}
|\Lambda_Q \rangle & \equiv & |\Lambda_Q; 0, {1 \over 2}, J_z \rangle \, 
\sim \Lambda_c , \,\,\,  \Lambda_b \, ,
\nonumber \\
|\Lambda_{Q1} \rangle & \equiv &|\Lambda_Q; 1, {1 \over 2}, J_z \rangle \,
\sim \Lambda_c (2593), \,\,\,  \Lambda_b (?) \, ,
\nonumber \\
|\Lambda_{Q1}^{*} \rangle & \equiv &|\Lambda_Q; 1, {3 \over 2}, J_z \rangle \,
\sim \Lambda_c (2625), \,\,\, \Lambda_b (?) \, .
\label{states}
\end{eqnarray}

The states of the doublet are degenerate in the combined limit up to NNLO.
This leads to an order $\lambda$ ambiguity in the determination of the energy 
of the first excited state. One way to define this energy is to take the
spin-averaged mass of the doublet:
\begin{equation} 
m_{\bar{\Lambda}_{Q}^{*}}  \equiv {1\over 3} m_{\Lambda_{Q1}} + 
{2\over 3} m_{\Lambda_{Q1}^{*}} \, .
\label{BSAM}
\end{equation}
In the charm sector the mass $m_{\bar{\Lambda}_{c}^{*}}$ is 
approximately $2610 \, MeV$. Similarly, the heavy meson mass, $m_H$,
can be taken as the spin-averaged mass of the corresponding pseudoscalar and
pseudovector mesons:
\begin{eqnarray} 
m_{\bar{D}} & \equiv & {1\over 4} m_D + {3\over 4} m_{D^*} \approx 1980 \, MeV
\, , \nonumber \\
m_{\bar{B}} & \equiv & {1\over 4} m_B + {3\over 4} m_{B^*} \approx 5310 \, MeV
\, .
\label{MSAM}
\end{eqnarray}

The excitation energies up to NLO are given by,
\begin{equation}
\Delta m_c \equiv m_{\bar{\Lambda}_{c}^{*}} - m_{\Lambda_c}  = 
\sqrt{\kappa \over \mu_c} + {5 \over 4!}{\alpha \over \kappa \mu_c}+
{\cal O}(\lambda^{3/2}) \, ,
\label{deltamc} 
\end{equation}
\begin{equation}
\Delta m_b \equiv m_{\bar{\Lambda}_{b}^{*}} - m_{\Lambda_b}  = 
\sqrt{\kappa \over \mu_b} + {5 \over 4!}{\alpha \over \kappa \mu_b}+
{\cal O}(\lambda^{3/2}) \, ,
\label{deltamb}
\end{equation}
where the order $\lambda$ corrections are determined by treating the 
$\alpha\, x^4/4!$ term in Hamiltonian perturbatively. As indicated above,
these excitation energies are completely determined in terms of two constants,
$\kappa$ and $\alpha$.

In addition to the spectroscopic observables, the effective theory predicts
the dominant semileptonic form factors and electromagnetic decay rates
of the excited heavy baryons. Up to NLO these observables are completely
determined in terms of $\kappa$ and $\alpha$. Semileptonic form factors and
electromagnetic decays have yet to be measured. 

As discussed in Ref.~\cite{hb3}, the effective theory in the combined limit
can reliably predict the dominant semileptonic form factors only for
the transfer velocities, $|\delta \vec v|$, of order $\lambda^{3/4}$. However,
this is the range of the velocity transfers within which the form factors
significantly change from their values at zero recoil. 

In the combined limit the hadronic amplitude of the $\Lambda_b \to \Lambda_c 
\ell \bar{\nu}$ decay up to NLO is given in terms of a single independent
form factor \cite{hb3}:
\begin{equation}
\langle \Lambda_c (\vec{v}^\prime)|\bar{c}\gamma^{\mu}
(1-\gamma_5) b|\Lambda_b (\vec{v}) 
\rangle = \Theta (z) \, 
\bar{u}_c (\vec{v}^\prime) \gamma^{\mu}(1-\gamma_5) u_b (\vec{v})
\left (1 +{\cal O}(\lambda) \right) \, ,
\label{Theta} 
\end{equation}
where Dirac spinors are normalized by 
$\bar{u}_{Q} (\vec{v}, s) u_{Q}(\vec{v}, s)= 1$. 

The form factor $\Theta (z)$ in eq.~(\ref{Theta}) is given as a function
of the kinematic variable $z$ (instead of the more usual velocity transfer 
$|\delta \vec v|$ or momentum transfer $q$) defined by:
\begin{equation}
z \equiv {m_{N}|\delta \vec v| \over (\sqrt{\mu_b}+\sqrt{\mu_c})^{1/2}} \,. 
\label{z}
\end{equation}
As a function of the velocity transfer $|\delta \vec v|$ the form factor
$\Theta$ has an essential singularity in the combined limit, while
$\Theta (z)$ is a smooth function. The new kinematic variable $z$ is easily
related to the more familiar variable---the velocity transfer parameter
$w=v\cdot v^{\prime}$ where $v$ and $v^{\prime}$ are the 4-velocities
of the initial and final states:
\begin{equation}
z \equiv {m_{N}\sqrt{2(w -1)} \over (\sqrt{\mu_b}+\sqrt{\mu_c})^{1/2}}
\left(1+{\cal O}(\lambda^{3/2}) \right) \,. 
\label{zw}
\end{equation}
The relation in eq.~(\ref{zw}) is valid for velocity transfers of order 
$\lambda^{3/4}$.

The form factor $\Theta (z)$ was calculated up to NLO in Ref.~\cite{hb3}: 
\begin{eqnarray}
& \Theta (z)= {2 \sqrt{2}\,  \mu_{b}^{3/8} \mu_{c}^{3/8} \over 
(\sqrt{\mu_b}+\sqrt{\mu_c})^{3/2}} \exp \left(-z^2 \over 2\sqrt{\kappa}\right) 
\nonumber \\
& \left (1+{\alpha \over 4! \kappa^{5/2} (\sqrt{\mu_b}+\sqrt{\mu_c})} 
\left ({45 \kappa(\sqrt{\mu_b}-\sqrt{\mu_c})^{2} \over 16 \kappa^{3/2} 
\sqrt{\mu_b\mu_c}} +{5 \, z^2 \sqrt{\kappa}} -{1\over 4} z^4 \right ) \right ) 
\left(1+{\cal O}(\lambda^{3/2})\right) \, . &
\label{ThetaNLO}
\end{eqnarray}
Useful quantities extractable from experiment are the value of 
$\Theta (z)$ at zero recoil and the curvature at zero recoil (the slope of 
$\Theta (z)$ at $z=0$ vanishes). Up to NLO, these observables
are given by \cite{hb3}:
\begin{equation}
\Theta_0 \equiv \Theta (z=0)= {2 \sqrt{2} \,\, \mu_{b}^{3/8} \mu_{c}^{3/8} 
\over (\sqrt{\mu_b}+\sqrt{\mu_c})^{3/2}}
\left ( 1+{\alpha \over 4!} {45 (\sqrt{\mu_b}-\sqrt{\mu_c})^{2} \over 
16 \kappa^{3/2}\sqrt{\mu_b\mu_c} (\sqrt{\mu_b}+\sqrt{\mu_c})} \right ) 
\left(1+{\cal O}(\lambda^{3/2})\right)  \, ,
\label{Thetazero}
\end{equation}
\begin{equation}
\rho \equiv {\partial^2 \Theta \over \partial z^2} (z=0)=
- {2 \sqrt{2} \,\, \mu_{b}^{3/8} \mu_{c}^{3/8} \over 
\sqrt{\kappa} (\sqrt{\mu_b}+\sqrt{\mu_c})^{3/2}}
\left ( 1+{\alpha \over 4!} {45 (\sqrt{\mu_b}-\sqrt{\mu_c})^{2} -
160 \sqrt{\mu_b \mu_c} \over 16 \kappa^{3/2}
\sqrt{\mu_b\mu_c} (\sqrt{\mu_b}+\sqrt{\mu_c})} \right ) 
\left(1+{\cal O}(\lambda^{3/2})\right)\, .
\label{rho}
\end{equation}
At leading order, $\Theta_0$ is independent of $\kappa$ and $\alpha$; it is an
overall normalization which is approximately $0.998$. This value is very close
to the heavy quark effective theory (HQET) value of unity, despite the fact 
that the present expansion is not the pure heavy quark expansion. The 
curvature at zero recoil and a slope at zero recoil of the form factor 
$\Theta$ as a function of $w$ are related by,
\begin{equation}
{\partial \Theta \over
\partial w}(w=1)={m_{N}^{2} \over \sqrt{\mu_b}+\sqrt{\mu_c}}\,\, 
\rho \,.
\label{rhoomega}
\end{equation}
Note that $\rho$ is of order $\lambda^0$ while $\partial \Theta /\partial 
w \,(w=1)$ is of order $\lambda^{3/2}$ in the combined limit.

There are two semileptonic decay channels of $\Lambda_b$ corresponding to the 
decays to the doublet of 
the first excited state: $\Lambda_b \to \Lambda_{c1} \ell \bar{\nu}$ and
$\Lambda_b \to \Lambda_{c1}^{*}\ell \bar{\nu}$. As shown in Ref.~\cite{hb3},
the dominant form factor which determines the hadronic amplitudes for
both of these channels is given by,
\begin{eqnarray}
\langle \Lambda_{c1} (\vec{v}^\prime)|\bar c \gamma^\mu (1-\gamma_5) b |
\Lambda_b (\vec{v}) \rangle & =& \sqrt{3}\, \Xi (z)
\bar{u}_{c} (\vec{v}^\prime) \gamma^{\mu}(1-\gamma_5)u_b (\vec{v})
\left( 1+{\cal O}(\lambda)\right ) \, ,
\nonumber \\
\langle \Lambda_{c1}^{*} (\vec{v}^\prime)|\bar c \gamma^\mu (1-\gamma_5) b |
\Lambda_b (\vec{v}) \rangle & = & \Xi (z) \, 
\bar{u}_{c\nu} (\vec{v}^\prime) (\sigma^{\nu\mu}\gamma_5 - g^{\mu\nu})
u_b (\vec{v}) \left( 1+{\cal O}(\lambda)\right ) \, ,
\label{Ksi} 
\end{eqnarray}
where the Rarita-Schwinger spinors are normalized by
$\bar{u}_{\nu}(\vec{v}, s) u^{\nu}(\vec{v}, s)=-1$. 

For the velocity transfers of order $\lambda^{3/4}$ the form factor $\Xi (z)$
is determined in Ref.~\cite{hb3} up to NLO:
\begin{eqnarray}
& \Xi \, (z) = {4 \, z \,  \mu_{b}^{3/8} \mu_{c}^{5/8} \over 
\kappa^{1/4}(\sqrt{\mu_b}+\sqrt{\mu_c})^{2}}  
\exp\left( -{z^2 \over 2 \sqrt{\kappa}} \right)&
\nonumber \\
& \left ( 1+{\alpha \over 4! \kappa^{5/2} (\sqrt{\mu_b}+\sqrt{\mu_c})}
\left ({(105 \mu_b - 230 \sqrt{\mu_b \mu_c}+ 45 \mu_c) \kappa \over 16 
\sqrt{\mu_b \mu_c}} + {13\over2}\, \kappa^{1/2} z^2 -{1\over 4}\, z^4\right ) 
\right ) \left(1+ {\cal O}(\lambda^{3/2}) \right )\, . &
\label{KsiNLO}
\end{eqnarray} 
At zero recoil the form factor $\Xi(z)$ vanishes. The slope of $\Xi (z)$ at
zero recoil is given by,
\begin{equation}
\sigma \equiv {\partial \Xi \over \partial z} \, (z=0)
= {4 \,\, \mu_{b}^{3/8} \mu_{c}^{5/8} \over 
\kappa^{1/4}(\sqrt{\mu_b}+\sqrt{\mu_c})^{2}}
\left ( 1+{\alpha \over 4!} 
{105 \mu_b - 230 \sqrt{\mu_b \mu_c}+ 45 \mu_c \over 16 \kappa^{3/2} 
\sqrt{\mu_b \mu_c} (\sqrt{\mu_b}+\sqrt{\mu_c})} \right )
\left (1+{\cal O}(\lambda^{3/2}) \right) \, .
\label{sigmaNLO}
\end{equation}  

Radiative decays of the excited heavy baryons might also be measured. Due to
the small available phase space, radiative decays are expected to have a large 
branching ratio for excited $\Lambda_c$ and $\Lambda_b$ baryons; moreover, 
they may be the dominant decay channels of the excited $\Lambda_b$ baryons. 
It was shown in a 
model-independent way in Ref.~\cite{hb3}, that in the combined limit the 
electromagnetic decays of the doublet ($\Lambda_{Q1}$, $\Lambda_{Q1}^{*}$)
in the charm and bottom sectors are dominated by dipole radiation. 
The total decay rate averaged over the initial state and summed over all final
states is the same for $\Lambda_{Q1}$ and $\Lambda_{Q1}^{*}$ \cite{hb3}. The
total decay rates in the charm and bottom sector up to NLO are found to be: 
\begin{eqnarray}
\Gamma(\Lambda_{c1} \to \Lambda_c \, \gamma) & = &
{1 \over 6 } e^2 \kappa {\left (m_{\bar{D}}-m_N \over m_{\bar{D}}m_N \right)^2}
\left  (1- {\alpha \over 4!} {5 \over \sqrt{\kappa^3 \mu_c}} \right )
\left (1+{\cal O}(\lambda) \right) \, ,
\label{Gammac} \\
\Gamma(\Lambda_{b1} \to \Lambda_b \, \gamma) & = &
{1\over 6 }e^2 \kappa {\left (m_{\bar{B}}+m_N \over m_{\bar{B}}m_N \right )^2}
\left (1- {\alpha \over 4!} {5 \over \sqrt{\kappa^3 \mu_b}}\right )
\left (1+{\cal O}(\lambda) \right) \, ,
\label{Gammab}
\end{eqnarray}  
where $e$ is the electromagnetic coupling constant ($e^2 \approx 1/137$).
It is interesting to note that the ratio of these decay rates at LO is 
$\Gamma(\Lambda_{c1}\to\Lambda_c\,\gamma)/
\Gamma(\Lambda_{b1} \to \Lambda_b \, \gamma) \approx 0.2$, which
significantly differs from $1$---the HQET prediction at leading order in 
$1/m_Q$. The deviation is due to the large recoil of the heavy quark against
the brown muck in the charm system where $m_{N}/m_{\bar{D}} \approx 1/2$.
  
At leading order, the values of the heavy baryon observables, 
eqs~(\ref{deltamc}), (\ref{deltamb}), (\ref{Thetazero}), (\ref{rho}), 
(\ref{sigmaNLO}), (\ref{Gammac}) and (\ref{Gammab}), depend on one 
parameter---$\kappa$. This 
parameter can be eliminated at LO in a model-independent way by using the mass
splitting between the ground state and the first orbitally excited state of 
$\Lambda_c$ in eq.~(\ref{deltamc}). At LO, the model independent predictions 
are given in Table.~\ref{table1}. The theoretical uncertainty of these 
model-independent predictions can be large; corrections are nominally of 
relative order $\lambda^{1/2} \sim (1/3)^{1/2} \sim 60 \%$. To improve 
accuracy, the next-to-leading order corrections must be included. The 
predictions at NLO contain theoretical uncertainty of relative order $\lambda$
which is nominally $\sim 1/3 \sim 30 \,\%$.
   
\begin{table}
\caption{Model-independent predictions at LO. The observables are defined in
eqs.~(\ref{deltamc}), (\ref{deltamb}), (\ref{Thetazero}), (\ref{rho}), 
(\ref{sigmaNLO}), (\ref{Gammac}) and (\ref{Gammab}). Corrections are of 
relative order $\lambda^{1/2}$.}
\begin{tabular}{|r|r|r|r|r|r|r|}
Fitted & \multicolumn{6}{c|}{Predictions} \\
Observable & $\Delta m_b \, (MeV)$ & $\Theta_0$ & $\rho \, (MeV^{-3/2})$ & 
$\sigma \, (MeV^{-3/4})$ & $\Gamma_c \, (MeV)$ & $\Gamma_b \, (MeV)$ \\ 
\hline
$ \Delta m_c \approx 330 \, MeV $ & $300$ & $\approx 1$ & $-1.2\times 
10^{-4}$ & 0.011 & 0.025 & 0.13 
\end{tabular}
\label{table1}
\end{table}

At NLO, heavy baryon observables, eqs.~(\ref{deltamc}), (\ref{deltamb}), 
(\ref{Thetazero}), (\ref{rho}), (\ref{sigmaNLO}), (\ref{Gammac}) and 
(\ref{Gammab}),  contain an additional constant---$\alpha$. Model-independent 
predictions at this order can, in principle, be made by eliminating $\kappa$ 
and $\alpha$ in terms of any pair of observables. In order to experimentally 
test these predictions apart from the presently known value of the
excitation energy of $\Lambda_c$ ($\Delta m_c \approx 330 \, MeV$), another 
observable ({\it eg.} a derivative of the semileptonic decay rate, $\rho$ or 
$\sigma$, or the total radiative decay rate of the first excited state of
$\Lambda_c$) has to be measured. The sensitivity of different 
observables to the value of $\alpha$ varies. One therefore has to  
judiciously choose an additional observable. The next-to-leading order 
contribution to such observables should be large enough numerically so as to 
be stable against the NNLO corrections. Based on the sensitivity to $\alpha$,  
we will argue that the best observables (in addition to $\Delta m_c$)
are either the second derivative of the form factor $\Theta (z)$ at zero 
recoil---$\rho$---or the slope of the form factor $\Xi (z)$ at zero 
recoil---$\sigma$---or the total radiative decay rate of the first excited
state of $\Lambda_c$ or $\Lambda_b$---$\Gamma_c$ or $\Gamma_b$.

To estimate the size of the NLO corrections we can rescale the constants
$\kappa$ and $\alpha$ and the observables in  eqs.~(\ref{deltamc}), 
(\ref{deltamb}), (\ref{Thetazero}), (\ref{rho}), (\ref{sigmaNLO}), 
(\ref{Gammac}) and (\ref{Gammab}) using the typical momentum scale of the
collective degrees of freedom. It is natural to define this scale by,
\begin{equation}
\Lambda\equiv(\mu_{c} {\Delta m_c}^2)^{1/3} \approx 410\, MeV \, ,
\label{scale}
\end{equation} 
with $\Lambda^3$ being equal to the value of $\kappa$ determined from the
leading order term in eq.~(\ref{deltamc}): $\kappa = \mu_{c} {\Delta m_c}^2
\left (1+{\cal O}(\lambda^{1/2} \right)$. In terms of the scale $\Lambda$,
dimensionless constants $\bar{\kappa}$ and $\bar{\alpha}$ can be defined by,
\begin{equation}
\bar{\kappa}\equiv \Lambda^{-3} \kappa \, , \, \,\,\,\,\,
\bar{\alpha}\equiv \Lambda^{-5} \alpha \, .
\label{barkapalph}
\end{equation}
The dimensionless constant $\bar{\kappa}$ is unity when only leading order 
terms in the combined expansion of the observables in eqs.~(\ref{deltamc}),
(\ref{deltamb}), (\ref{Thetazero}), (\ref{rho}), (\ref{sigmaNLO}), 
(\ref{Gammac}) and (\ref{Gammab}) are kept. When NLO terms are included, the 
constant $\bar{\kappa}$ has the form:
\begin{equation}
\bar{\kappa}=(1+\delta \bar{\kappa})\left (1 + {\cal O}(\lambda) \right) \, ,
\label{kappaNLO}
\end{equation}
where $\delta \bar{\kappa}$ is of relative order $\lambda^{1/2}$. Hence,
when working up to NLO, the LO terms containing $\bar{\kappa}$ may be 
linearized with respect to $\delta\bar{\kappa}$. Moreover, since in all of the 
observables in eqs.~(\ref{deltamc}), (\ref{deltamb}), (\ref{Thetazero}), 
(\ref{rho}), (\ref{sigmaNLO}), (\ref{Gammac}) and (\ref{Gammab}) the
constants $\kappa$ and $\alpha$ contribute only via the ratio $\alpha/\kappa^n$
(where $n$ is an integer or half-integer), the value of $\bar{\kappa}$ in
these expressions can be taken to be unity when NNLO corrections are neglected.
In addition to using dimensionless constants $\bar{\kappa}$ and
$\bar{\alpha}$, we can rescale the observables according to their dimensions
using appropriate powers of $\Lambda$. The corresponding dimensionless
observables are given by,
\begin{eqnarray}
\bar{\Delta m_c} & \equiv & \Delta m_c\, \Lambda^{-1} =
(1+{1\over2}\delta\bar{\kappa})
\sqrt{\Lambda^3 \over \mu_c} + {5 \over 4!}{\bar{\alpha}\Lambda^2
\over \mu_c} + {\cal O}(\lambda^{3/2})
\nonumber \\
&\approx&( 0.80 \, (1+{1\over2}\delta \bar{\kappa}) + 0.13 \, \bar{\alpha})
\left(1+{\cal O}(\lambda)\right) \, ,
\nonumber \\
\bar{\Delta m_b} & \equiv & \Delta m_b \, \Lambda^{-1}=
(1+{1\over2}\delta\bar{\kappa})
\sqrt{\Lambda^3 \over \mu_b} + {5 \over 4!}{\bar{\alpha}\Lambda^2
\over \mu_b} + {\cal O}(\lambda^{3/2})
\nonumber \\
&\approx& (0.72 \, (1+{1\over2}\delta \bar{\kappa}) +0.11 \, \bar{\alpha})
\left(1+{\cal O}(\lambda)\right) \, , 
\nonumber \\
\bar{\Theta}_0 & \equiv & \Theta_0 =
{2 \sqrt{2} \,\, \mu_{b}^{3/8} \mu_{c}^{3/8} \over 
(\sqrt{\mu_b}+\sqrt{\mu_c})^{3/2}}
\left ( 1+{\bar{\alpha}\sqrt{\Lambda} \over 4!}  
{45 (\sqrt{\mu_b}-\sqrt{\mu_c})^{2}
\over 16 \sqrt{\mu_b\mu_c} (\sqrt{\mu_b}+
\sqrt{\mu_c})} \right ) \left(1+{\cal O}(\lambda)\right)
\nonumber \\
&\approx & (0.998 +5.6 \times 10^{-4} \, \bar{\alpha})
\left(1+{\cal O}(\lambda)\right) \, , 
\nonumber \\
\bar{\rho} & \equiv & \rho\, \Lambda^{3/2} =
-{2 \sqrt{2} \,\, \mu_{b}^{3/8} \mu_{c}^{3/8}\over \Lambda^{3/2}(\sqrt{\mu_b}+
\sqrt{\mu_c})^{3/2}}
\left (1-{1\over2}\delta\bar{\kappa}+{\bar{\alpha} \sqrt{\Lambda}\over 4!} 
{45 (\sqrt{\mu_b}-\sqrt{\mu_c})^{2}-160 \sqrt{\mu_b \mu_c}\over 16 
\sqrt{\mu_b\mu_c} (\sqrt{\mu_b}+\sqrt{\mu_c})} \right ) 
\left(1+{\cal O}(\lambda)\right)
\nonumber \\
&\approx& (-0.998\, (1-{1\over2}\delta\bar{\kappa}) + 0.15 \, \bar{\alpha})
\left(1+{\cal O}(\lambda)\right)\, , 
\nonumber \\
\bar{\sigma} & \equiv & \sigma \, \Lambda^{3/4}= 
{4 \,\, \mu_{b}^{3/8} \mu_{c}^{5/8} \over \Lambda^{3/4}
(\sqrt{\mu_b}+\sqrt{\mu_c})^{2}}
\left(1-{1\over4}\delta\bar{\kappa}+{\bar{\alpha}\over 4!} 
{105 \mu_b - 230 \sqrt{\mu_b \mu_c}+ 45 \mu_c \over 16 \Lambda^{1/2}
\sqrt{\mu_b \mu_c} (\sqrt{\mu_b}+\sqrt{\mu_c})} \right )
\left (1+{\cal O}(\lambda) \right)
\nonumber \\
&\approx& (0.969 \, (1-{1\over4}\delta \bar{\kappa}) - 0.07 \, \bar{\alpha})
\left(1+{\cal O}(\lambda)\right)\, , 
\nonumber \\
\bar{\Gamma}_c & \equiv & {6\over e^2}\Gamma_c\, \Lambda^{-1}= 
{\left (m_{\bar{D}}-m_N \over m_{\bar{D}}m_N \right)^2} \Lambda^3
\left  (1+\delta\bar{\kappa}- {\bar{\alpha}\sqrt{\Lambda} \over 4!} 
{5 \over \sqrt{\mu_c}} \right )
\left (1+{\cal O}(\lambda) \right)
\nonumber \\
&\approx& (0.05 \, (1+\delta\bar{\kappa})-0.01 \, \bar{\alpha})
\left(1+{\cal O}(\lambda)\right)\, , 
\nonumber \\
\bar{\Gamma}_b & \equiv & {6\over e^2}\Gamma_b \,\Lambda^{-1} =
{\left (m_{\bar{B}}+m_N \over m_{\bar{B}}m_N \right )^2} \Lambda^3
\left (1+\delta\bar{\kappa} - {\bar{\alpha}\sqrt{\Lambda} \over 4!} 
{5 \over \sqrt{\mu_b}}\right )\left (1+{\cal O}(\lambda)
\right)
\nonumber \\
& \approx & (0.26 \, (1+\delta \bar{\kappa})-0.04 \, \bar{\alpha})
\left(1+{\cal O}(\lambda)\right) \, ,
\label{rescaled}
\end{eqnarray}
where we linearized the terms containing $\bar{\kappa}$. In the NLO terms this
amounts to taking $\bar{\kappa}$ to be $1$ as was discussed above. In the 
second set of equalities in eq.~(\ref{rescaled}) the experimental values for 
the masses have been used.

The rescaled observables in eq.~(\ref{rescaled}) depend linearly on the 
dimensionless constants $\delta\bar{\kappa}$ and $\bar{\alpha}$. The value of 
$\bar{\alpha}$ is expected to be of order unity. The NLO corrections for 
all observables in eq.~(\ref{rescaled}), except for $\Theta_0$, are  
$7\,\%$ to $20\,\%$ when $\bar{\alpha}$ is taken to be unity. The NLO 
contributions to the the value of $\Theta_0$ is suppressed by four orders of 
magnitude. The leading order value of $\Theta_0$ is very close to unity (see 
Table~\ref{table1})---the leading order value predicted by HQET. The 
deviation from unity is an exceptionally poor way to extract $\alpha$ since 
the sensitivity is so small numerically that one expects higher order terms to
be significant. The sensitivities with respect to $\bar{\kappa}$ can be used 
to decide which observables are best suited to extract values of $\bar{\kappa}$
and $\bar{\alpha}$.

Formally, the $1/m_Q$ and $1/N_c$ corrections in the combined expansion are of
the same order. However, the heavy quark expansion is generically expected to 
be more a accurate approximation for heavy baryons than the large $N_c$ 
expansion. If $\Delta m_b$ is used apart from $\Delta m_c$ to eliminate 
$\kappa$ and $\alpha$, the resulting  relations contain mass differences, 
$\Delta m_c-\Delta m_b$. In the heavy quark limit this difference is zero up to
$1/m_Q$ corrections. At leading order in the combined expansion this
difference is approximately $30 \, MeV$ (Table~\ref{table1}). The NNLO
corrections which are beyond the accuracy we are working to can be estimated
from the mass splitting of the states of the doublet ($\Lambda_{c1}$,
$\Lambda_{c1}^{*}$). This splitting---approximately $40 \, MeV$---is the NNLO
effect containing $1/m_Q$ and $1/N_c$ corrections. Hence, predictions
containing $\Delta m_c-\Delta m_b$ and $\Delta m_c+\Delta m_b$ are sensitive
to NNLO corrections and should not be used to extract $\kappa$ and $\alpha$.

Any of the remaining observables---$\rho$, $\sigma$ and $\Gamma_c$ 
(or $\Gamma_b$)---can be used along with $\Delta m_c$ to eliminate $\kappa$ 
and $\alpha$. Using $\bar{\Delta m_c}$ and $\bar{\rho}$ as primary observables,
we can obtain model-independent predictions for other observables. These
predictions are listed in the second column in Table~\ref{table2}. Similarly,
using eq.~(\ref{rescaled}), $\bar{\kappa}$ and $\bar{\alpha}$ can be 
extracted from pairs ($\bar{\Delta m_c}$, $\bar{\sigma}$) and ($\Delta m_c$,
$\bar{\Gamma_c}$). Corresponding model-independent relations are
given in the third and the fourth columns in Table~\ref{table2}. The 
corrections to relations in Table~\ref{table2} are generically of order
$\lambda$ which nominally is $1/3 \sim 30 \, \%$. In order to test NLO 
predictions, measurements of one of the observables---the curvature of the
form factor $\Theta (z)$ at zero recoil, the slope of the form factor 
$\Xi (z)$ at zero recoil, or the total radiative decay rate of the first
excited state of $\Lambda_c$---has to be measured.   

\begin{table}
\caption{Model-independent predictions at NLO. Observables are given in
 eqs.~(\ref{deltamc}), (\ref{deltamb}), (\ref{Thetazero}), (\ref{rho}), 
(\ref{sigmaNLO}), (\ref{Gammac}) and (\ref{Gammab}). The typical scale 
$\Lambda$ is defined as $\Lambda \equiv (\mu_c {\Delta m_c}^2)^{1/3} 
\approx 410 \, MeV$. Corrections are of relative order $\lambda$.}
\begin{tabular}{|r|r|r|r|}
& \multicolumn{3}{c|}{Fitted Observables} \\
Predictions & $\Delta m_c$ and $\rho$ & $\Delta m_c$ and $\sigma$ & 
$\Delta m_c$ and $\Gamma_c$ \\
\hline
$\Delta m_b \, \Lambda^{-1} $ & $1.29+0.57 \,(\rho \Lambda^{3/2})$ & 
$0.77-0.05\, (\sigma \Lambda^{3/4})$ & $0.71+0.27\,(\Gamma_c \Lambda^{-1})$ \\
\hline
$\Theta_0$ & $0.95-0.05 \,(\rho \Lambda^{3/2})$ & 
$0.99+0.004\,(\sigma \Lambda^{3/4})$ &
$0.999-0.02\, (\Gamma_c \Lambda^{-1})$ \\
\hline
$\rho \, \Lambda^{3/2}$ & & $-0.92-0.08 \, (\sigma \Lambda^{3/4})$ &  
$-1.01+0.46 \,(\Gamma_c \Lambda^{-1})$ \\
\hline
$\sigma \, \Lambda^{3/4}$ & $0.25-0.72 \,(\rho \Lambda^{3/2})$ & &
$0.98-0.33\,(\Gamma_c \Lambda^{-1})$ \\
\hline
$\Gamma_c \, \Lambda^{-1}$ & $(2.67+2.62\,(\rho \Lambda^{3/2}))\times 10^{-3}$ 
& $(2.67-2.15 \, (\sigma \Lambda^{3/4}))\times 10^{-4}$ & \\
\hline
$\Gamma_b \, \Lambda^{-1}$ & $(1.27+1.24 \,(\rho \Lambda^{3/2}))\times 10^{-2}$
& $(1.30-1.02\,(\sigma \Lambda^{3/4}))\times 10^{-3}$ & 
$(0.029+5.77\,(\Gamma_c \Lambda^{-1}))\times 10^{-3}$
\end{tabular}
\label{table2}
\end{table}

We have discussed the model-independent relations between observables which 
arise in the combined heavy quark and large $N_c$ expansion at leading and 
next-to-leading order. These observables were derived in the framework of
an effective theory in the combined heavy quark and large $N_c$ limit in 
\cite{hb2,hb3}. At leading order in the combined expansion a model independent
expressions for a number of heavy baryon observables contain one constant.
This coefficient can be eliminated using the known value of the excitation
energy of the first excited state of $\Lambda_c$, $\Delta m_c$. The resulting
predictions up to corrections of order $\lambda^{1/2}$ are listed in 
Table~\ref{table1}. To obtain predictions at next-to-leading order in the
combined expansion, an additional observable is required. At the present time
the data is not available. Observables in  eqs.~(\ref{deltamc}), 
(\ref{deltamb}), (\ref{Thetazero}), (\ref{rho}), (\ref{sigmaNLO}), 
(\ref{Gammac}) have different sensitivities with respect to the 
next-to-next-to-leading order corrections. We have discussed observables 
(when data becomes available) should be used
in addition to the excitation energy $\Delta m_c$ to predict the
remaining observables. The model-independent relations at NLO are given in
Table~\ref{table2}.

\acknowledgements
This work is supported by the U.S.~Department of Energy grant 
DE-FG02-93ER-40762.

\end{document}